\documentclass[12pt]{article}
\usepackage{epsfig,amssymb,latexsym}

\newcommand{\bq}{\begin{equation}}
\newcommand{\eq}{\end{equation}}
\newcommand{\bqa}{\begin{eqnarray}}
\newcommand{\eqa}{\end{eqnarray}}
\newcommand{\ben}{\begin{enumerate}}
\newcommand{\een}{\end{enumerate}}
\newcommand{\bc}{\begin{center}}
\newcommand{\ec}{\end{center}}
\newcommand{\bqb}{\begin{eqnarray*}}
\newcommand{\eqb}{\end{eqnarray*}}

\def\pr#1#2#3{ Phys. Rev. ${\bf{#1}}$:#2 (#3)}
\def\prl#1#2#3{ Phys. Rev. Lett. ${\bf{#1}}$:#2 (#3)}
\def\pl#1#2#3{ Phys. Lett. ${\bf{#1}}$:#2 (#3)}

\def\np#1#2#3{ Nucl. Phys. ${\bf{#1}}$:#2 (#3)}
\def\zp#1#2#3{ Z. f. Phys. ${\bf{#1}}$:#2 (#3)}

\def\epj#1#2#3{ Eur. Phys. J. ${\bf{#1}}$:#2 (#3)}
\def\ijmp#1#2#3{ Int. J. Mod. Phys. ${\bf{#1}}$:#2 (#3)}

\def\cpc#1#2#3{Comput. Phys. Commun. ${\bf{#1}}$:#2 (#3)}

\def\aop#1#2#3{Annals of Phys. ${\bf{#1}}$:#2 (#3)}

\def\polon#1#2#3{Acta Phys. Polon. ${\bf{#1}}$:#2 (#3) }

\def\eg{{\it e.g. }}

\def\etal{{\it et.al. }}

\def\mz{m_Z}
\def\mzd{m_Z^2}

\begin{document}
\pagenumbering{arabic}
\thispagestyle{empty}
\def\thefootnote{\fnsymbol{footnote}}
\setcounter{footnote}{1}

\begin{flushright}
PTA/PTA/06-48 \\
hep-ph/0610085 \\
October 2006\\
Revised version\\
 \end{flushright}
\vspace{2cm}
%---------------------title ---------------------------------------
\begin{center}
{\Large\bf A FORTRAN code for
$\gamma \gamma \to Z Z $ in SM and MSSM\footnote{Work
co-funded by the European Union - European Social fund and the
National fund PYTHAGORAS – EPEAEK II} }  \\
%-----------------------------------------------------------------
{\large Th. Diakonidis$^a$, G.J. Gounaris$^a$ and J. Layssac$^b$ }\\
\vspace{0.2cm}
$^a$Department of Theoretical Physics, Aristotle
University of Thessaloniki,\\
Gr-54006, Thessaloniki, Greece.\\
\vspace{0.2cm}
$^b$Laboratoire de Physique
Th\'{e}orique et Astroparticules, UMR 5207,\\
Universit\'{e} Montpellier II,
 F-34095 Montpellier Cedex 5.\\
\vspace{0.2cm}

 \vspace*{1cm}

{\bf Abstract}

\end{center}

Through  the present paper, the code {\bf gamgamZZ} is presented,
 which may be used
to calculate all possible  observables related to the process $\gamma \gamma \to ZZ$,
in either the Standard Model (SM), or  the minimal sypersymmetric
standard model (MSSM) with real parameters.

\def\thefootnote{\arabic{footnote}}
\setcounter{footnote}{0}
\clearpage

\section{Introduction}

The starting of the LHC operation in 2007 brings  closer  the moment where
we will be able to see whether the simplest Standard Model (SM) higgs
particle really exists \cite{Higgs-WG},
and whether  the much searched for scale of New Physics (NP)
 lies in the TeV range \cite{LHC-ILC}.

Among the many such NP forms, the most widely  studied is certainly
${\cal N}=1 $ Supersymmetry (SUSY) in four dimensions. Hints supporting  the
minimal supersymmetric model (MSSM), with real electroweak scale parameters,
are supplied by the unification of the gauge couplings  at the scale of
$\sim 2\cdot 10^{16}$ GeV and the proximity of this scale to  the one needed
for understanding  the neutrino masses and leptogenesis \cite{neutrinos}.
The easy  accommodation of Dark Matter in R-parity conserving SUSY models,
is  also  encouraging.

But, if  SUSY candidates are discovered  at LHC  \cite{ILC-physics, SPA, light-h0},
the need to check in detail their  properties in an  international
linear collider (ILC) will be overwhelming \cite{TESLAee, CLICee, Heuer}.

This note concerns   the important option of ILC
to run as a $\gamma \gamma$
collider $ILC_{\gamma\gamma}$, using back scattering of $e^\pm$-beams by laser
photons \cite{TESLAgg, CLICgg, Telnov}. In many instances,
such a $\gamma \gamma$ collider is more efficient for
studying new physics, than the standard
$e^-e^+$-ILC  \cite{Zerwas-gamma, ggWW, Nactmann, Sekaric}.

A very important set of processes observable  in $ILC_{\gamma \gamma}$,
consists of  the neutral gauge boson production
\bq
\gamma \gamma \to \gamma \gamma,~ \gamma Z,~ ZZ ~~, \label{gauge-production}
\eq
which  are first realized at the 1-loop level, in both SM and MSSM,
and could therefore  be very
sensitive to possible new physics.
Their analytical 1-loop study was first performed
for   SM in \cite{gg-theory, gZ-theory,
ZZ-theory}, and subsequently  extended to include
all possible  MSSM contributions, with real $\mu$ and
soft SUSY parameters \cite{gggg, gggZ, ggZZ1, ggZZ2}.

It has been found there that, at c.m. energies above 200GeV, the dominant
helicity amplitudes  for the processes  (\ref{gauge-production}) are mostly imaginary
and satisfy helicity conservation (HC), which forces
the sum of the helicities of the two incoming photons
to be equal to the sum of the helicities
of the two outgoing gauge bosons \cite{HC}. This
striking phenomenon mainly comes from the $W$-loop,
and could  make these processes  a very interesting field of study
in $ILC_{\gamma \gamma}$.
Thus, if  \eg   SUSY candidates are  discovered at LHC,
the study of (\ref{gauge-production})
at the contemplated 1\% level of accuracy, should provide
important  checks of their nature.\\

In  studying  (\ref{gauge-production}),
the independent helicity amplitudes
have been expressed in terms of the simplest
 Veltman-Passarino functions $B_0,C_0,D_0$ \cite{Veltman}.
The resulting formulae are  quite manageable for the $\gamma \gamma $ and
$\gamma Z$  cases.
For  $\gamma \gamma \to Z Z$  though,
the analytic expressions of \cite{ggZZ1, ggZZ2}, particularly for  MSSM,
  are so complicated, that  a public numerical  code will be very
  useful.

  The purpose of the present paper is to release  such code  "gamgamZZ",
      applying  to  SM and MSSM,
  for any set of real  parameters \cite{code}.
  For both models,  the code calculates
the independent helicity amplitudes  for $\gamma \gamma \to Z Z$,
as well as  the  observable cross sections  in $LC_{\gamma\gamma}$,
for the case where     the polarizations of the final $ZZ$-pair are summed over.
The  definitions of these  amplitudes and cross sections
are given in Section 2.
In Section 3, we discuss the code and give some
numerical examples. Section 4 contains the Conclusions.

\section{The $\gamma \gamma \to Z Z$ helicity amplitudes and cross sections}

We are interested in  the process
\bq
\gamma (p_1,\lambda_1) \gamma (p_2,\lambda_2) \to
Z (p_3,\lambda_3) Z (p_4,\lambda_4) \ \ ,
\label{ggZZ-process}
\eq
where the momenta $p_j$ and helicities $\lambda_j$,
of the incoming and outgoing particles
are indicated in parentheses. The generic form of the relevant 1-loop
 Feynman diagrams is  indicated in Fig.\ref{diag-fig}a,b.
The  corresponding helicity amplitudes,
denoted as\footnote{Their sign is related to the sign of
the $S$-matrix through   $S_{\lambda_1 \lambda_2
\lambda_3\lambda_4}= 1+i (2\pi)^4 \delta(p_4+p_3-p_1-p_2)
F_{\lambda_1 \lambda_2 \lambda_3\lambda_4}$. }
$F_{\lambda_1 \lambda_2 \lambda_3\lambda_4}(\beta_Z, t, u)$,
are expressed in terns of the kinematical variables
\bqa
 && s\equiv s_{\gamma \gamma}
 =(p_1+p_2)^2 = \frac{4\mzd}{1-\beta_Z^2} ~~, \label{kin-beta} \\
&&  t=(p_1-p_3)^2=\mzd-\frac{s}{2}(1-\beta_Z\cos\theta)  ~ ~,~ ~ \label{kin-t} \\
&& u=(p_1-p_4)^2 =\mzd-\frac{s}{2}(1 +\beta_Z\cos\theta)~ ,  \label{kin-u}
\eqa
where $\theta$ is the c.m scattering angle, and
      $\beta_Z$  coincides with the $Z$-velocity in the $ZZ$ c.m. frame,
      provided  its positive sign is selected from (\ref{kin-beta}). The possible
 values of $(\lambda_1,\lambda_2)$ are $\pm 1$; while those of
 $(\lambda_3,\lambda_4)$ are$(\pm1,0)$.

Bose statistics, combined with the Jacob-Wick
(JW) phase conventions \cite{JW-convention},
demand
\bqa
F_{\lambda_1 \lambda_2 \lambda_3\lambda_4}(\beta_Z, t,u) &=&
F_{\lambda_2 \lambda_1 \lambda_4\lambda_3}(\beta_Z, t, u)
(-1)^{\lambda_3-\lambda_4} \ , \label{Bose2} \\
F_{\lambda_1 \lambda_2 \lambda_3\lambda_4}(\beta_Z, t,u) &=&
F_{\lambda_2 \lambda_1 \lambda_3\lambda_4}(\beta_Z, u,t)
(-1)^{\lambda_3-\lambda_4} \ , \label{Bose1} \\
F_{\lambda_1 \lambda_2 \lambda_3\lambda_4}(\beta_Z,t,u) &=&
F_{\lambda_1 \lambda_2 \lambda_4\lambda_3}(\beta_Z,u,t)
 \ , \label{Bose12}
\eqa
while CP invariance  implies
\bq
F_{\lambda_1 \lambda_2 \lambda_3\lambda_4}(\beta_Z , t, u) =
F_{-\lambda_1,-\lambda_2,- \lambda_3,-\lambda_4}(\beta_Z, t, u)
(-1)^{\lambda_3-\lambda_4} \  .  \label{parity}
\eq
These relations  allow the calculation of the complete set of the
36 helicity amplitudes, in terms
of the 10 basic ones
\bqa
&& F_{+++-}(\beta_Z, t,u)~ , ~ F_{++++}(\beta_Z, t,u), \nonumber \\
& &   F_{+-++}(\beta_Z, t,u)~, ~ F_{+-00}(\beta_Z, t,u)~, \nonumber \\
 &&F_{++00}(\beta_Z, t,u)~ ,~  F_{+++0}(\beta_Z, t,u), \nonumber \\
 & & F_{+-+0}(\beta_Z, t,u)~ , ~  F_{+-+-}(\beta_Z, t,u )~, \nonumber \\
 && F_{++--}(\beta_Z, t,u) ~,~ F_{++-0}(\beta_Z, t,u) ~.
\label{10basic}
\eqa

In \cite{ggZZ1, ggZZ2} these amplitudes have been analytically expressed in terms
of the $B_0,~ C_0,~ D_0 $ Passarino-Veltman \cite{Veltman} functions.
The code presented bellow is based on these expressions.
 As far  as   SM is concerned,  these results agree
with those   of \cite{ZZ-theory},
except for a trivial misprint in $F_{+-+0}$,  pointed out in
\cite{ggZZ2}.
\\

As expected on the basis of the helicity conservation (HC) theorem, established in
\cite{HC} for any two body process,
the dominant helicity amplitudes in either  SM or MSSM,  should obey
\bq
\lambda_1+\lambda_2 =\lambda_3+\lambda_4 ~~, \label{HCrule}
\eq
for    $s$ and $|t|$ much larger than all squared masses in the model.
  In MSSM, HC is  an exact asymptotic theorem; while
  in SM, it  has only been proven to 1-loop leading logarithmic accuracy
 \cite{HC}.

This means  that, at asymptotic energies and finite angles,
 all  MSSM amplitudes of the set  (\ref{10basic}) should
 vanish, except  $(F_{++++},~ F_{+-+-},~ F_{+-00})$,
 which increase  logarithmically
 with energy and are mostly imaginary \cite{ggZZ1} .

 We have checked numerically that the  1-loop results of
\cite{ggZZ1, ggZZ2}, indeed obey this rule.
This is particularly striking  for   $F_{++00}$,
where various  contributions,
often involving ratios of gauge, chargino and charged higgs masses,
have been  seen to cancel each other exactly at asymptotic $s,|t|$.
The validity of HC provides a very nice check of our   code.

 HC is essentially true also  for SM,
 the only difference being that the amplitudes that violate
(\ref{HCrule}) go asymptotically to  small generally non-vanishing  constants.\\

Backscattering polarized laser photons  off polarized $e^\mp$-beams,
 would realize   an $ILC_{\gamma \gamma}$ collider
 \cite{TESLAgg, Telnov}. Restricting to the case that the final
 $Z$ polarizations are not studied,
the observable cross sections at $ILC_{\gamma \gamma}$
 are given by \cite{ggZZ1, ggZZ2}
\bqa
{d \sigma_0(\gamma \gamma \to ZZ) \over d\cos\theta}&=&
\left ({\beta_Z\over 128 \pi\hat{s}}\right )
\sum_{\lambda_3\lambda_4} [|F_{++\lambda_3\lambda_4}|^2
+|F_{+-\lambda_3\lambda_4}|^2] ~ ,  \label{sig0} \\
{d {\sigma}_{22}(\gamma \gamma \to ZZ)\over d\cos\theta} &=&
\left ({\beta_Z\over 128 \pi\hat{s}}\right )\sum_{\lambda_3\lambda_4}
[|F_{++\lambda_3\lambda_4}|^2
-|F_{+-\lambda_3\lambda_4}|^2]  \ , \label{sig22} \\
{d {\sigma}_{3}(\gamma \gamma \to ZZ)\over d\cos\theta}&=&
\left ({-\beta_Z\over 64 \pi\hat{s}}\right ) \sum_{\lambda_3\lambda_4}
Re[F_{++\lambda_3\lambda_4}F^*_{-+\lambda_3\lambda_4}]  \ ,
\label{sig3} \\
{d  \sigma_{33}(\gamma \gamma \to ZZ) \over d\cos\theta}& = &
\left ({\beta_Z\over 128\pi\hat{s}}\right ) \sum_{\lambda_3\lambda_4}
Re[F_{+-\lambda_3\lambda_4}F^*_{-+\lambda_3\lambda_4}] \ ,
\label{sig33} \\
{d {\sigma}^\prime_{33}(\gamma \gamma \to ZZ)\over d\cos\theta}&=&
\left ({\beta_Z\over 128 \pi\hat{s}}\right ) \sum_{\lambda_3\lambda_4}
Re[F_{++\lambda_3\lambda_4}F^*_{--\lambda_3\lambda_4}] \  ,
\label{sig33prime} \\
{d {\sigma}_{23}(\gamma \gamma \to ZZ)\over d\cos\theta}& = &
\left ({\beta_Z\over 64 \pi\hat{s}}\right ) \sum_{\lambda_3\lambda_4}
Im[F_{++\lambda_3\lambda_4}F^*_{+-\lambda_3\lambda_4}] ~, \label{sig23}
\eqa
in terms of   the helicity
amplitudes  satisfying  (\ref{Bose2}-\ref{10basic}).

\section{The gamgamZZ code}

The released code gamgamZZ.tar.gz \cite{code},
consists of four FORTRAN codes. Two of them, {\bf sm1\_1} and {\bf mssm1\_1},
calculate the  10 basic $\gamma\gamma \to ZZ$ helicity amplitudes (\ref{10basic}),
in  SM and  MSSM respectively.
The remaining helicity amplitudes may then be obtained using
(\ref{Bose2}-\ref{parity}). The needed higgs width may be calculated
from \eg \cite{SUSY-HIT}.

For compiling these codes,  LoopTools \cite{looptools} is required.
All input  parameters are contained in input files  called
"susypar.in" in both, the SM and the MSSM case.
All dimensional parameters are  in TeV.
The results for each $\gamma\gamma$ c.m. energy, are given for
$N_\theta+1$  equal steps of $\cos\theta $, starting from $\cos(\theta_a)$
and ending in $\cos(\theta_b)$. The integer parameter $N_\theta$,
and the angles $\theta_a$ and $\theta_b$ in  degrees, are specified in the
in-file.

As an example we give in Table 1 the results of the {\bf sm1\_1} output, for
5 of the 10 helicity  amplitudes listed in (\ref{10basic}) and $N_\theta=8$ .
A similar output is created for the remaining five basic amplitudes.
The c.m. energy is taken at  $\sqrt{s}=0.5$TeV. The exact
form of all outputs is explained in the Readme.dat file accompanying the codes.

\begin{table}[htb]
\begin{center}
{ Table 1: Form of the {\bf sm1\_1} output
for 5 of the 10 helicity amplitudes listed in (\ref{10basic}).
A similar form  appears in the same output for the description
of the remaining five amplitudes of (\ref{10basic}).}\\
\vspace*{0.3cm}
    \hspace*{-0.7cm}
\begin{tiny}
\begin{tabular}{||c|c|c|c|c|c|c|c|c|c|c||}
\hline \hline
$\cos\theta$ & $ReF_{++++}$ & $ImF_{++++}$ & $ReF_{+-++}$ & $ImF_{+-++}$&
$ReF_{+-00}$ & $ImF_{+-00}$ & $ReF_{++00}$ & $ImF_{++00}$&
$ReF_{+++0}$ & $ImF_{+++0}$    \\ \hline
    0.866 &   0.144E-1  &   0.162 &   0.671E-3 &   0.179E-2 &   -0.328E-3  &  0.360E-2 &  -0.267E-2 &  -0.473E-2  &  0.551E-3  &  0.183E-2 \\
    0.707 &   0.104E-1  &   0.137 &  0.655E-3  &  0.204E-2  &  -0.561E-3   & 0.633E-2  & -0.246E-2  & -0.426E-2   & 0.351E-3  &  0.124E-2  \\
    0.500 &  0.772E-2 &   0.120 &   0.472E-3 &   0.228E-2  & -0.832E-3  &  0.857E-2 &  -0.227E-2  & -0.390E-2  &  0.164E-3  &  0.783E-3  \\
    0.259 &   0.631E-2  &  0.110  &  0.322E-3  &  0.244E-2 &  -0.104E-2  &  1.000E-2 &  -0.215E-2 &  -0.368E-2  &  0.594E-4  &  0.381E-3 \\
   0.0 &   0.588E-2  &  0.107  &  0.2679E-3  &  0.250E-2 &  -0.112E-2  &  0.105E-1 &  -0.211E-2  & -0.361E-2  &  0.0  & 0.0 \\
   -0.259 &   0.631E-2  &  0.110  &  0.322E-3  &  0.244E-2 &  -0.104E-2  &  1.000E-2 &  -0.215E-2 &  -0.368E-2  & -0.594E-4  & -0.381E-3 \\
   -0.500  &   0.771E-2  &  0.120  &  0.472E-3  &  0.228E-2 &  -0.832E-3  &  0.857E-2 &  -0.227E-2 &  -0.390E-2  & -0.164E-3  & -0.783E-3 \\
   -0.707  &  0.104E-1  &  0.137  &  0.655E-3  &  0.204E-2 &  -0.561E-3  &  0.633E-2 &  -0.246E-2 &  -0.426E-2  & -0.351E-3  & -0.124E-2 \\
   -0.866  &  0.1443E-1  &  0.162  &  0.671E-3  &  0.179E-2 &  -0.328E-3  &  0.360E-2 &  -0.267E-2 &  -0.473E-2  & -0.551E-3  & -0.183E-2 \\
 \hline \hline
\end{tabular}
 \end{tiny}
\end{center}
\end{table}

The other two codes,  {\bf sm2} and {\bf mssm2}  are compiled similarly
and  calculate
the differential cross sections
(\ref{sig0}-\ref{sig23}) in SM and MSSM respectively.
They follow a similar format,
fully  explained in the aforementioned  Readme.dat file.
These results do not of course include the  $ILC_{\gamma\gamma}$luminosity
functions. But it    should be easy to combine them with a code
like CompAZ in order to calculate ZZ production in a
  realistic $ILC_{\gamma\gamma}$ environment \cite{CompAZ}.

\begin{figure}[p]
\vspace*{3cm}
\[
\epsfig{file=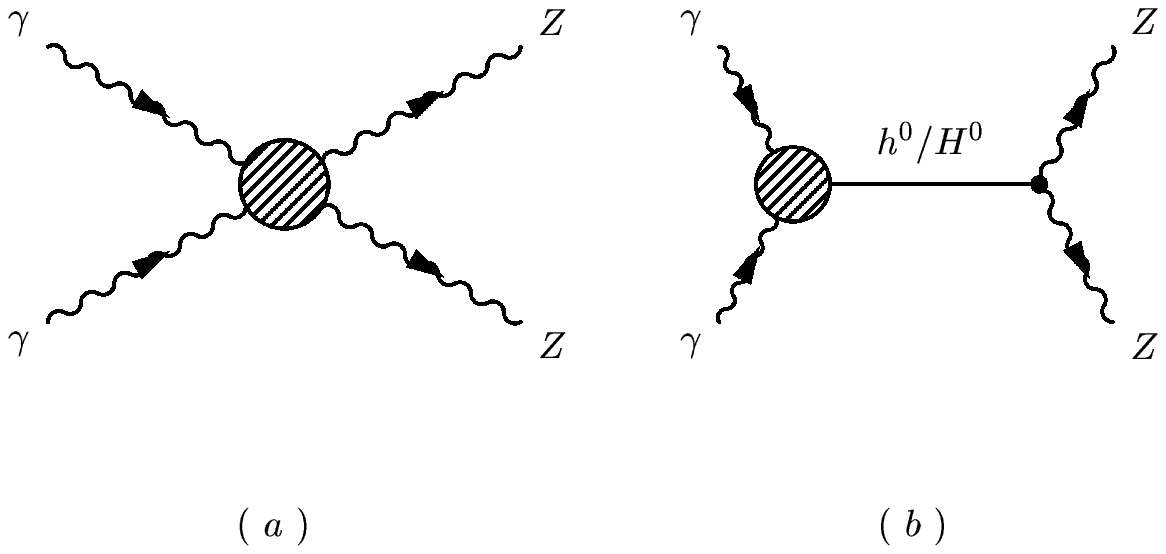,height=5.cm}
\]
%\vspace*{0.5cm}
\caption[1]{Feynman Diagrams for the $\gamma \gamma \to ZZ$
process in SM and MSSM models.}
\label{diag-fig}
\end{figure}

The  {\bf sm1\_1}, {\bf sm2} codes give  the exact  SM 1-loop
contribution from the diagrams in Fig.\ref{diag-fig}a,b.

Correspondingly,  the MSSM codes {\bf mssm1\_1}, {\bf mssm2}
contain  the exact    $W$, higgs, quark and lepton loop
contributions due to the box and triangular diagrams
in Fig.\ref{diag-fig}a,b.
The  sfermion and chargino 1-loop triangular
contributions from the diagrams\footnote{The chargino
or sfermion mixing cannot induce a non-diagonal
$Z$-coupling contribution in this case.} in Fig.\ref{diag-fig}b
are also calculated exactly.

In  these codes though, we have neglected
the sfermion and chargino mixing effects, arising
from the non-diagonal $Z$ couplings,
in the box diagrams of Fig.\ref{diag-fig}a.
This is always a very good approximation
for the (very small) sfermion box contribution.
It  is also true for  the chargino box contribution
 for all helicity amplitudes, except
for the amplitudes    involving two longitudinal
 $Z$ bosons.
 The reason for this is because
 the  high energy chargino-box contribution to the production of a longitudinal
 gauge boson is determined by
 the  gaugino-higgsino-higgs coupling \cite{equivalence},
for which   chargino mixing cannot be neglected.
The only  amplitudes of (\ref{10basic}) belonging to
this category are  $F_{++00}$ and $F_{+-00}$,  for which the
chargino   box contribution is approximated
by the asymptotic part of (A.37, A.38) of \cite{ggZZ1} and
 (A.57, A.58, A.63, A.64 ) of \cite{ggZZ2}.
 This is an  adequate approximation,
 since these  contributions are rather  energy independent
 and very small.

 If it becomes necessary,  the exact
 chargino  1-loop box contribution  could  be included
  in our codes, by
 using  the full  analytical expressions appearing
 in (A.36-A.66) of  the Appendix of \cite{ggZZ2}. \\

It is also important to note that when
 using LoopTools \cite{looptools} for
calculating   the needed  Passarino-Veltman functions,
 problems may appear related
to the small masses of the quarks and leptons of the
first two families.  We have checked that if the LoopTools1.2 version
is used, there is no difficulty.
For  LoopTools2 and newer versions though, the masses of the quarks and charge leptons
of the first two families should be increased
to   around   7 or 8 MeV, to overcome this problem.
This does not reduce the accuracy
    of our results, and  is  easily  done by changing the
input parameters in the in-file.\\

\begin{figure}[p]
\vspace*{-2.0cm}
\[
\hspace{-1.2cm}\epsfig{file=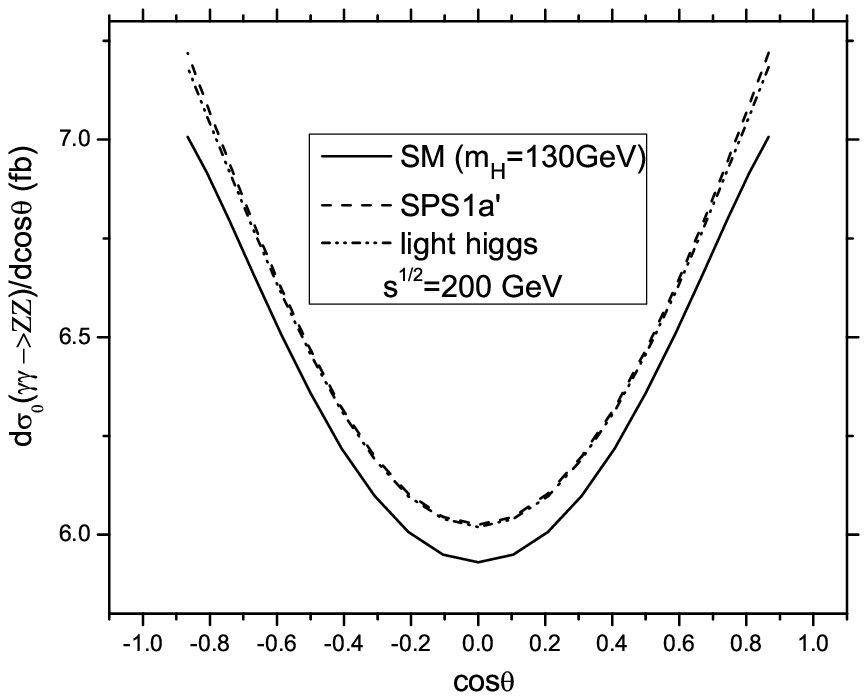,height=8.cm}
\]
%\vspace*{-1.cm}
\[
\hspace{-1.2cm} \epsfig{file=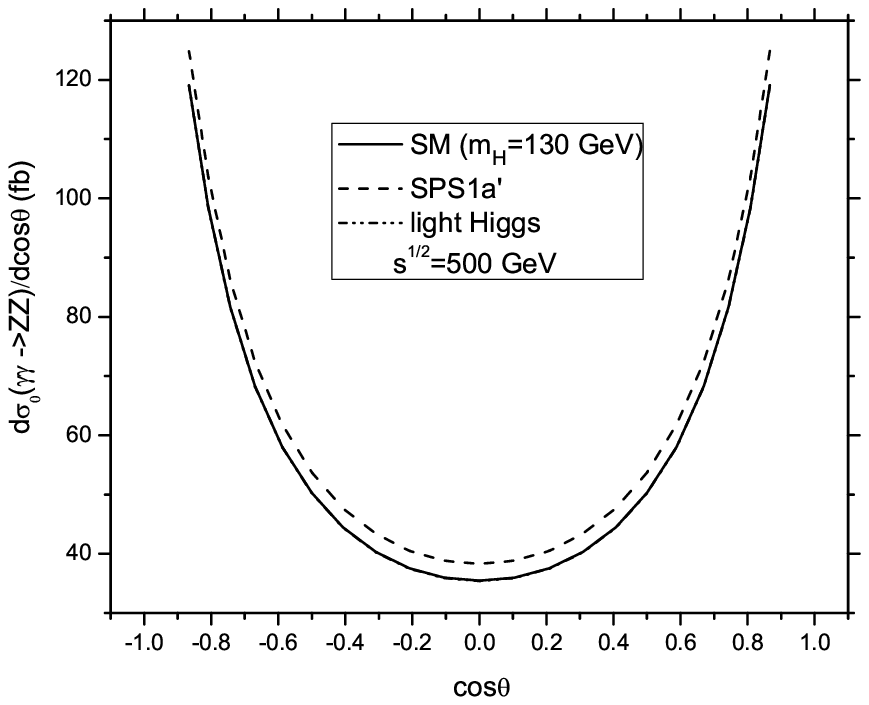,height=8.cm}
\]
%
%\vspace*{-1.cm}
\caption[1]{Differential cross section
$ d\sigma_0(\gamma \gamma \to ZZ) / d \cos \theta $ as a function of $\cos\theta$,
in SM,  $SPS1a^\prime$ \cite{SPA} and the light higgs \cite{light-h0}
benchmark model.}
\label{dsigma0}
\end{figure}

\begin{figure}[p]
\vspace*{-2.0cm}
\[
\hspace{-1.2cm}\epsfig{file=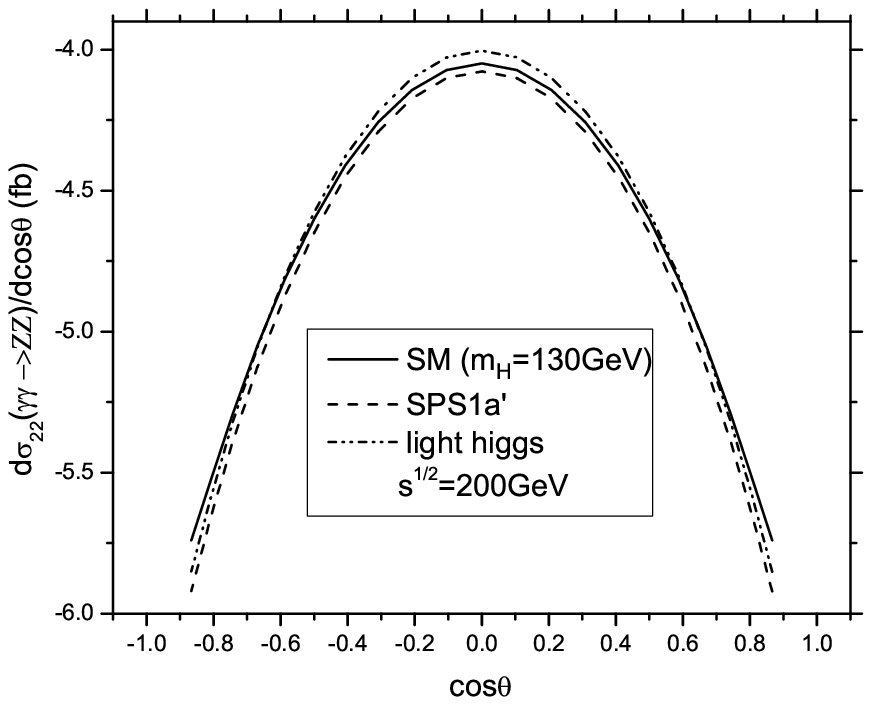,height=8.cm}
\]
%\vspace*{-1.cm}
\[
\hspace{-1.2cm} \epsfig{file=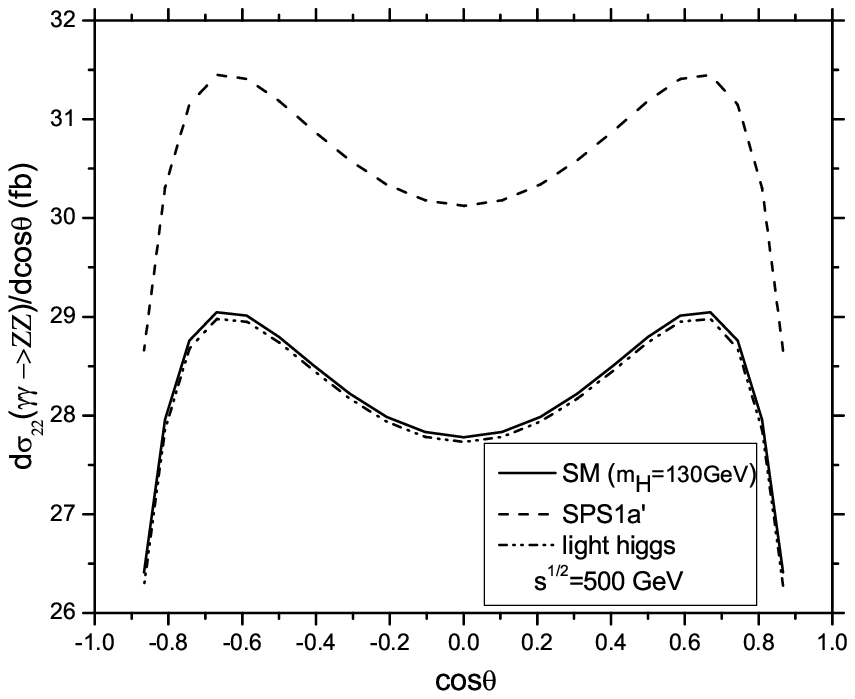,height=8.cm}
\]
%
%\vspace*{-1.cm}
\caption[1]{Differential cross section
$ d\sigma_{22}(\gamma \gamma \to ZZ) / d \cos \theta $ as a function of $\cos\theta$,
in SM,  $SPS1a^\prime$ \cite{SPA} and the light higgs \cite{light-h0}
benchmark model.}
\label{dsigma22}
\end{figure}

\begin{figure}[p]
\vspace*{-2.0cm}
\[
\hspace{-1.2cm}\epsfig{file=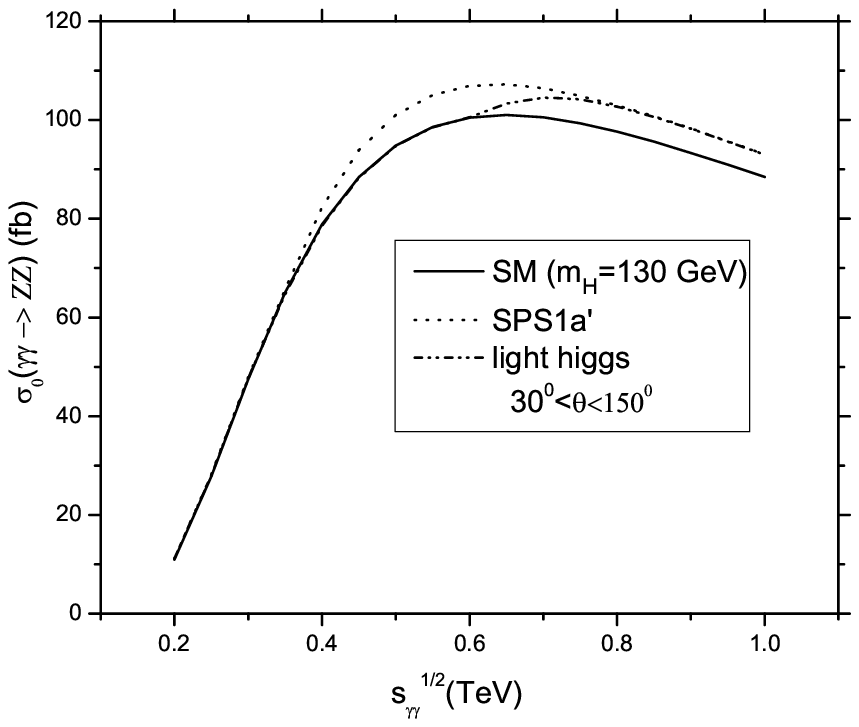,height=9.cm}
\]
%\vspace*{-1.cm}
\[
\hspace{-1.2cm} \epsfig{file=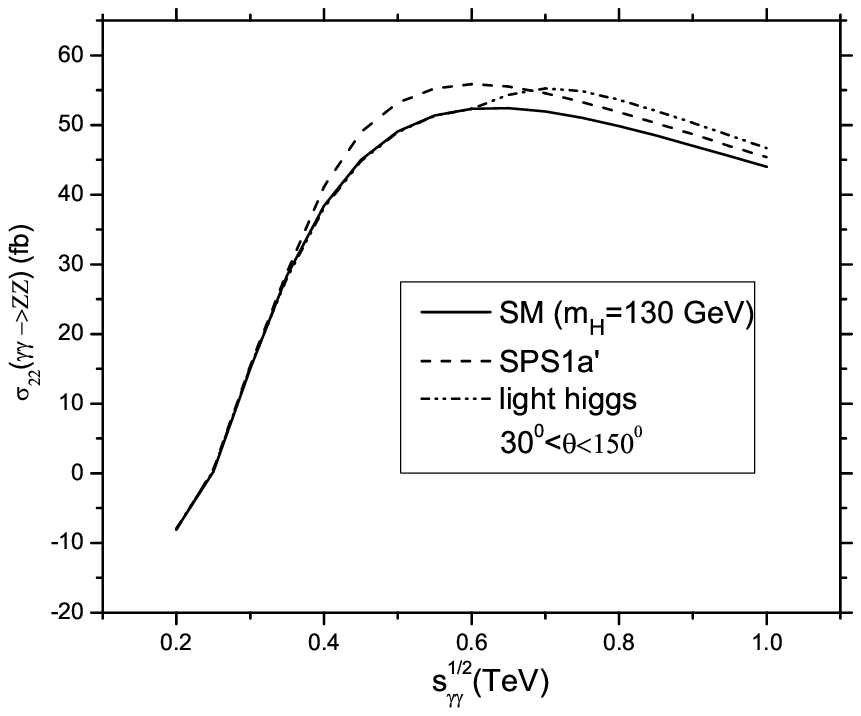,height=9.cm}
\]
%
%\vspace*{-1.cm}
\caption[1]{Integrated cross sections
$ \sigma_{0}(\gamma \gamma \to ZZ)$ and
$ \sigma_{22}(\gamma \gamma \to ZZ)$,
as  functions of the center of mass photon-photon energy
$\sqrt{s}=\sqrt{s_{\gamma\gamma}}$, in SM,
$SPS1a^\prime$ \cite{SPA} and the light higgs \cite{light-h0} benchmark model.}
\label{sigma0-sigma22}
\end{figure}

As an example,  we present
in Figs.\ref{dsigma0}-\ref{sigma0-sigma22} the differential
and integrated cross sections $\sigma_0$ and $\sigma_{22}$,
defined in (\ref{sig0}, \ref{sig22}), for SM (with $m_H=130 {\rm GeV}$)  and
for the two benchmark MSSM models $SPS1a'$ \cite{SPA} and
 "light higgs"   \cite{light-h0}.
$SPS1a'$ is an mSUGRA model with $(m_0=70 ~{\rm GeV}, ~m_{1/2}=250 ~{\rm GeV},
~A_0=-300 ~{\rm GeV},~\tan\beta=10)$;
while in "light higgs",  the electroweak scale parameters
have been selected as $M_1=150$GeV,
$M_2=300$GeV, $M_3=600$GeV, $A_t=A_b=A_\tau=700$GeV,
$\mu=700$GeV, $m_{A^0}=104$ GeV, $\tan\beta=34.4$
and all sfermion masses are set at 340GeV \cite{light-h0}.
In all cases, the top mass was put at  $m_t=173.8 {\rm GeV}$.

The specific  cross sections $\sigma_0,~\sigma_{22}$ have been chosen because
they often give the largest effect. They are also the
ones that receive contributions from
the s-channel\footnote{In fact the only cross sections that receive such contributions
are  $\sigma_0,~\sigma_{22}, ~\sigma_{23}$; compare (\ref{sig0}-\ref{sig23}).}
higgs exchange diagrams of Fig.\ref{diag-fig}b.
Thus, if the SM higgs is above the $ZZ$-threshold but not too heavy,  a higgs peak
should be visible  in the $\sigma_0$ and $\sigma_{22}$-plots \cite{ggZZ2}.
This  applies also to MSSM, provided that
the heavier neutral higgs particle $H^0$
is   above the $ZZ$-threshold and  its couplings are appropriate;
see examples in  \cite{ggZZ2}.
Among the scenarios  considered here though, only   $SPS1a'$, implying
$m_{H^0}=0.424 {\rm TeV}$, satisfies the $2\mz$ constraint.
But the $H^0$ contribution   is so weak in this model,
that no  peak is visible in Figs.\ref{sigma0-sigma22}.

In any case, the SUSY modifications
to the SM predictions for $\sigma_0$ and $\sigma_{22}$,
 are generally  energy dependent
and sometimes considerably larger than $1\%$. Therefore, SUSY effects
in an $ILC_{\gamma \gamma}$ study of   $\gamma \gamma \to ZZ$, may
be observable and useful.\\

Before finishing this Section,  it is important    to remember that nowadays,
and surely much more in the future, computer codes  like
FORMCALC \cite{formcalc}  or  GRACE-LOOP
\cite{grace-loop} are being developed
which, starting from a given Lagrangian,
construct automatically all vertices\footnote{Possibly
employing useful  approximations,
like neglecting chargino mixing in boxes, as we have done here.},
draw the necessary   diagrams and calculate all amplitudes
in an analytic (or semi-analytic)
way. Applying to $\gamma \gamma \to ZZ$, this means that all work contained in \eg
\cite{ggZZ1, ggZZ2} is done in an automatic way, obviously deriving similar
analytic results. These analytic results should    then  be
transformed to a FORTRAN code, with no  need of extra labor.
Under the same choice of approximations,
this FORTRAN code should be similar to the one presented here, admittedly after
considerable labor. In any case though, no other such $\gamma \gamma \to ZZ$
code seems to exist at present.

\section{Conclusions}

We have  presented here the code gamgamZZ, contained in
 gamgamZZ.tar.gz  \cite{code}. This consists of four codes.
Codes  {\bf sm1\_1} and {\bf mssm1\_1}
 calculate the 10 basic helicity amplitudes (\ref{10basic}) for
$\gamma \gamma \to ZZ$, in   SM and  MSSM respectively;
while codes {\bf sm2} and {\bf mssm2} calculate the  cross sections
in (\ref{sig0}-\ref{sig23}).
When the $Z$ polarization is not looked at,
these cross sections  constitute
all possible observables in an $ILC_{\gamma \gamma}$.

The accompanying  Readme.dat file contains all necessary
instructions for compiling and running the codes \cite{code}.
The only restriction, applying to the MSSM case, is that the soft SUSY
breaking parameters and $\mu$ must be real. But if interest arises,
this restriction could be easily  overcome,
by  changing certain couplings in the codes.

Since the codes  {\bf sm1\_1} and {\bf mssm1\_1} determine all possible helicity
amplitudes, the results should also be useful for calculating
any kind of   $Z$-polarization effects observable in $ILC_{\gamma \gamma}$,
in either SM or MSSM.

\vspace*{1cm}
\underline{Acknowledgments}:\\
We are grateful to Fernand Renard and A. F. Zarnecki
for discussions.
 GJG  also gratefully acknowledges  the  support from the
 European Union program    MRTN-CT-2004-503369.

\newpage

\newpage


\begin{thebibliography}{99}

\bibitem{Higgs-WG} K.A. Assamagan \etal, hep-ph/0406152.
%
\bibitem{LHC-ILC} B.C. Allanach \etal,
 Les Houches physics at TeV colliders 2005 beyond the standard
 model working group: Summary report to the
 Les Houches Workshop on Physics at TeV Colliders,
 Les Houches, France, 2-20 May 2005, hep-ph/0602198
%
\bibitem{neutrinos} See \eg A. Strumia and F. Vissani, hep-ph/0606054.
%
\bibitem{ILC-physics} E.A. Baltz, M. Battaglia, M.E. Peskin and
T. Wizansky, hep-ph/0602187;
K. M\"{o}nig, \ijmp{A21}{1974}{2006}, hep-ph/0509159;
W. Killian and P.M. Zerwas,
Talk at the 2nd ILC Accelerator Workshop,
Snowmass, Colorado, 14-27 Aug 2005, hep-ph/06012217.
%
\bibitem{SPA} J.A. Aguilar-Saavedra \etal, (SPA convention),
\epj{C46}{43}{2006}, hep-ph/0511344; A Djouadi, M. Drees and Jean-Loic Kneur,
hep-ph/0602001; A. Djouadi, hep-ph/0503173.
%
\bibitem{light-h0}A. Belyaev \etal, heph/0609079.
%
\bibitem{TESLAee}``TESLA Technical Design Report Part I: Executive Summary,''
eds.\ F.~Richard, J.~R.~Schneider, D.~Trines and A.~Wagner,
DESY, Hamburg, 2001 [hep-ph/0106314]. \\
``TESLA Technical Design Report Part III: Physics at an $e^+e^-$
Linear Collider,''  eds.\ R.-D.~Heuer, D.~Miller, F.~Richard,
P.~M.~Zerwas, DESY, Hamburg, 2001 [hep-ph/0106315].
%
\bibitem{CLICee}J.~R.~Ellis, E.~Keil and G.~Rolandi,
``Options for Future Colliders at CERN,''
CERN-EP-98-03;\\
J.~P.~Delahaye \etal,
``CLIC---a Two-Beam Multi-TeV $e^+ e^-$ Linear Collider,''
in: {\it Proc.\ of the 20th Intl.\ Linac Conference LINAC 2000 }
ed.\ Alexander W.~Chao, eConf {C000821}, MO201 (2000), physics/0008064;\\
E.~Accomando {\it et al.}  [CLIC Physics Working Group Collaboration],
``Physics at the CLIC multi-TeV linear collider,''
CERN, Geneva, 2004, hep-ph/0412251.
%
\bibitem{Heuer}R.-D. Heuer, \polon{B37}{1039}{2006}.
%
\bibitem{TESLAgg} I.F. Ginzburg, G.L. Kotkin, V.G. Serbo
and V.I. Telnov, Nucl. Instr. and Meth. {\bf 205},  47 (1983);
I.F. Ginzburg, G.L. Kotkin, V.G. Serbo,
S.L. Panfil and V.I. Telnov,
Nucl. Instr. and Meth. {\bf 219},5 (1984);
J.H. K\"{u}hn, E.Mirkes
and J. Steegborn, \zp{C57}{615}{1993};
V. Telnov, hep-ex/0003024, hep-ex/0001029,
hep-ex/9802003, hep-ex/9805002, hep-ex/9908005;
I.F. Ginzburg, hep-ph/9907549;  R. Brinkman  hep-ex/9707017;
D.S. Gorbunov, V.A. Illyn, V.I. Telnov, hep-ph/0012175;
"TESLA Technical Design Report, Part VI, Chapter 1: Photon Collider at
TESLA", B.~Badelek {\it et al.}, DESY, Hamburg, 2001, [hep-ex/0108012].
%
\bibitem{CLICgg}
H.~Burkhardt and V.~Telnov,
"CLIC 3-TeV Photon Collider Option", CERN-SL-2002-013-AP.
%
\bibitem{Telnov} V.I. Telnov, talk at the PHOTON2005 and PLC2005 Conference,
Warsaw and Kazimierz, Poland, Sep 2005, \polon{B37}{1049}{2006},
Archive: physics/0604108.
%
\bibitem{Zerwas-gamma} M. M\"{u}hlleitner and P.M. Zerwas,
\polon{B37}{1021}{2006}, hep-ph/0511339
%
\bibitem{ggWW} G.J. Gounaris, J. Layssac  and
 F.M. Renard, \zp{C62}{139}{1994},  hep-ph/9309324;
 M. Bailargeon, G. B\/{e}langer and F. Boudjema, \np{B500}{1997}{224}, hep-ph/9701372;
 O. Nactman, F. Nagel, M. Pospischil, \epj{C45}{679}{2006}, hep-ph/0508132;
 O. Nachtmann, F. Nagel, M. Pospischil, A. Utermann \epj{C46}{93}{2006},
 hep-ph/0508133.
%
\bibitem{Nactmann} O. Nachtmann, F. Nagel, M. Pospischil,
A. Utermann, \epj{C45}{679}{2006}, hep-ph/0508132;
 \epj{C46}{93}{2006}, hep-ph/0508133.
%
\bibitem{Sekaric} J. Sekaric, DESY-THESIS-2005-042, hep-ph/0512307;
P. Nie\.{z}urawski, hep-ph/0503295.
%
\bibitem{gg-theory}G. Jikia and A. Tkabladze, \pl{B323}{453}{1994};
M. B\"{o}hm, R. Schuster \zp{C63}{219}{1994}.
%
\bibitem{gZ-theory}G. Jikia and A. Tkabladze, \pl{B332}{441}{1994}.
%
\bibitem{ZZ-theory}  G. Jikia \np{B405}{24}{1993}, \pl{B298}{224}{1993}.
%
\bibitem{gggg} G.J. Gounaris, P.I. Porfyriadis, F.M.
Renard, hep-ph/9812378, \pl{B452}{76}{1999},
\pl{B464}{350}{1999} (E); G.J. Gounaris, P.I. Porfyriadis, F.M.
Renard,  hep-ph/9902230, \epj{C9}{673}{1999}.
%
\bibitem{gggZ}G.J. Gounaris, J. Layssac, P.I. Porfyriadis and
 F.M. Renard,  hep-ph/9904450, \epj{C10}{499}{1999}.
%
\bibitem{ggZZ1}G.J. Gounaris, J. Layssac, P.I. Porfyriadis and
 F.M. Renard, \epj{C13}{79}{2000},  hep-ph/9909243.
%
\bibitem{ggZZ2}G.J. Gounaris, P.I. Porfyriadis and
 F.M. Renard, \epj{C19}{57}{2001},  hep-ph/0010006.
%
\bibitem{HC} G.J. Gounaris and F.M. Renard, \prl{94}{131601}{2005},
hep-ph/0501046, and the addendum in \pr{D73}{097301}{2006},
hep-ph/0604041.
%
\bibitem{Veltman} G. Passarino and M. Veltman \np{B160}{151}{1979}.
%
\bibitem{code} All neccessary files are contained in
gamgamZZ.tar.gz, which  may be downloaded from
http://users.auth.gr/$\sim$gounaris/FORTRANcodes/ .
%
\bibitem{JW-convention} M. Jacob and G.C. Wick,
\aop{7}{404}{1959}.
%
\bibitem{SUSY-HIT} A. Djouadi, M.M. M\"{u}hlleitner and M. Spira,
hep-ph/0609292;
A. Djouadi, J. Kalinowski and M. Spira,
HDECAY, hep-ph/9704448, \cpc{108}{56}{1998}.
%
\bibitem{looptools} T. Hahn, LoopTools,
http://www.feynarts.de/looptools/;
T. Hahn and M. P\'{e}rez-Victoria, hep-ph/9807565;
G.J. van Oldenborgh and J.A.M. Vermaseren,
\zp{C46}{425}{1990}.
%
\bibitem{equivalence} J.M. Cornwall, D.N. Levin, and G. Tiktopoulos,
\pr{D10}{1145}{1974};
M.S. Chanowitz, and M.K. Gaillard, \np{B261}{379}{1985};
G.J. Gounaris, R. K\"ogerler and H. Neufeld, \pr{D34}{3257}{1986}.
%
\bibitem{CompAZ} A. F. Zarnecki, \polon{B34}{2741}{2003},  hep-ex/0207021.
%
\bibitem{formcalc}T. Hahn and M. Perez-Victoria, \cpc{118}{153}{1999};
T. Hahn, hep-ph/0406288, hep-ph/0506201.
%
\bibitem{grace-loop}B. B\'{e}langer,
F. Boudjema, J. Fujimoto, T. Ishikawa, T. Kameko, K. Kato and Y. Shimizu,
hep-ph/0308080.
%


\end{thebibliography}
\end{document}